\documentclass[aps,pra,superscriptaddress,showkeys,showpacs,nofootinbib,twocolumn]{revtex4-1}
\usepackage{amsmath}    % need for subequations
\usepackage{graphicx}   % need for figures
\usepackage{array}
\usepackage[dvipdfx,cmyk,natbib]{xcolor}     % use if color is used in text
\usepackage[caption=false]{subfig}  % use for side-by-side figures
\usepackage[colorlinks,linkcolor=red,citecolor=brown,urlcolor=blue]{hyperref}
\usepackage{dsfont,bm}
\usepackage{amssymb}

\newcommand{\be}{\begin{equation}}
\newcommand{\ee}{\end{equation}}
\newcommand{\ben}{\begin{eqnarray}}
\newcommand{\een}{\end{eqnarray}}
\newcommand{\bes}{\begin{subequations}}
\newcommand{\ees}{\end{subequations}}
\newcommand{\bF}{\begin{figure}}
\newcommand{\eF}{\end{figure}}

\newcommand{\ttt}[1]{\text{#1}}
\newcommand{\bra}[1]{\langle{#1}\vert}
\newcommand{\ket}[1]{\vert{#1}\rangle}
\newcommand{\proj}[1]{\mbox{$|#1\rangle \!\langle #1 |$}}
\newcommand{\avg}[1]{\langle {#1} \rangle}

\begin{document}

\title{Simulating arbitrary Gaussian circuits with linear optics}

\author{L. Chakhmakhchyan}
\affiliation{Centre for Quantum Information and Communication, Ecole polytechnique de Bruxelles,
CP 165, Universit\'{e} libre de Bruxelles, 1050 Brussels, Belgium}
\affiliation{Quantum Engineering Technology Labs, H. H. Wills Physics Laboratory and Department of Electrical and Electronic Engineering, University of Bristol, BS8 1FD, Bristol, United Kingdom}
\author{N. J. Cerf}
\affiliation{Centre for Quantum Information and Communication, Ecole polytechnique de Bruxelles,
CP 165, Universit\'{e} libre de Bruxelles, 1050 Brussels, Belgium}
\date{\today}
\begin{abstract}
Linear canonical transformations of bosonic modes correspond to Gaussian unitaries, which comprise passive linear-optical transformations as effected by a multiport passive interferometer and active Bogoliubov transformations as effected by a nonlinear amplification medium. As a consequence of the Bloch-Messiah theorem, any Gaussian unitary can be decomposed into a passive interferometer followed by a layer of single-mode squeezers and another passive interferometer. Here, it is shown how to circumvent the need for active transformations. Namely, we provide a technique to simulate sampling from the joint input and output distributions of any Gaussian circuit with passive interferometry only, provided two-mode squeezed vacuum states are available as a prior resource. At the heart of the procedure, we exploit the fact that a beam splitter under partial time reversal simulates a two-mode squeezer, which gives access to an arbitrary Gaussian circuit without any nonlinear optical medium. This yields, in particular, a procedure for simulating with linear optics an extended boson sampling experiment, where photons jointly propagate through an arbitrary multimode Gaussian circuit, followed by the detection of output photon patterns.
\end{abstract}

\maketitle

\section{Introduction}
Recent advances in the theory and technology of quantum photonics have established it as one of the most promising candidate platforms to realize operational quantum technologies~\cite{revs}. The growing interest towards photonic architectures was originally triggered by the seminal protocol of Knill, Laflamme and Milburn (KLM) ~\cite{klm}, which demonstrates that universal quantum computation is possible using only passive linear optics components (i.e., beam splitters and phase shifters), single photon sources, photo-detectors and adaptive measurements upon ancillary resources. More recently, another wave  of remarkable progress in quantum photonics came, in part, after Aaronson and Arkhipov proved that highly demanding measurement-induced circuit control is not necessary to outperform a classical computer, i.e., to achieve the regime of a quantum advantage~\cite{BS}. Namely, sampling from the probability distribution of detecting single photons at the output of a linear-optical circuit, a task known as boson sampling, represents a problem that is intractable for a classical computer (see, e.g., Ref.~\cite{exp3} for its small-scale realizations). 

Although the boson sampling paradigm, as opposed to universal photonic quantum computing, does not require measurement-induced non-linearities, ancillary modes, nor post-selection, it still faces major challenges. In particular, entering the regime with quantum advantage would necessitate $\sim$50 photons distributed among $\sim$2500 modes~\cite{class, class1}, whereas the record is 5 photons in 16 modes~\cite{pan}. Moreover, such a linear optical device, despite being of fundamental importance, suffers from a lack of practical applications. For instance, an application for calculating molecular spectra is currently available~\cite{molec}, which has recently motivated proof-of-principle demonstrations~\cite{molec_exp, molec_exp2}. 

A possible strategy to overcome this state of affairs is to develop specialized sub-universal photonic setups, which lie in-between linear-optics and universal quantum computation. That is, identify a class of photonic circuits augmented with post-processing so to implement a restricted set of non-linearities. Here, we adopt this very approach and develop an optical scheme enabling us to simulate sampling from an arbitrary Gaussian circuit, i.e., any Gaussian unitary acting on bosonic modes. More precisely, our simulation provides a method for sampling from the joint input and output distributions of an arbitrary Gaussian circuit where the inputs are photon number states (with a specified probability distribution) and the outputs result from photon counting. For the sake of simplicity, we refer to this method as ``simulation of a Gaussian circuit". Importantly, Gaussian transformations have arisen to a privileged status in continuous-variable quantum information (where bosonic modes play the role of qubits, while Gaussian gates replace Clifford qubit gates), and have proven to be of a great practical interest in quantum computation, simulation, communication, as well as metrology~\cite{molec, rev_gaus, GBS, tensor, ising, metrology} (we briefly recall Gaussian states and transformations in Appendix~A). 

Our method for simulating Gaussian circuits relies on the Bloch-Messiah decomposition~\cite{BM} but introduces a major improvement. This decomposition implies that an arbitrary Gaussian transformation can always be mapped onto two linear-optical circuits intermitted by a layer of single-mode squeezers, hence requiring non-linear optical media. In contrast, our approach circumvents the need for in-line nonlinearity and requires two-mode squeezed vacuum states as a prior resource only. The building block of our procedure lies in that a two-mode squeezer is equivalent to a beam splitter undergoing partial time reversal \cite{cewqo}, allowing the conversion between passive and active optics (the degree of squeezing can be chosen arbitrarily, simply by tuning the beam-splitter transmissivity). Then, by making use of time symmetry considerations similar to those leading to the twofold version of scattershot boson sampling \cite{twofold, scattershot}, supplemented with a random-walk sampling algorithm for data-processing, we eventually construct a linear-optical simulator of the Bloch-Messiah decomposition. Our setup therefore reveals a special class of linear-optical circuits augmented with post-processing, which can simulate any Gaussian circuit.
%It may thus be viewed as the continuous-variable Gaussian analogue of the universal KLM scheme.
Furthermore, current photonic technologies, including integrated light sources, on-chip photo-detectors, and programmable circuitry should make the implementation of our scheme feasible~\cite{source, detect, circuit, exp1, detect, stefano, integrated}.

\section{Beam splitter under partial time reversal}

In the usual, predictive approach of quantum mechanics, one deals with the preparation of a system followed by its time evolution, and ultimately its measurement. The probability of the measurement outcome conditional on the preparation variable is given by Born's rule. In the retrodictive approach of quantum mechanics \cite{aharonov1964}, one post-selects the instances where a particular measurement outcome was observed and considers the probability of the preparation variable conditional on this outcome. The reverse Born's rule is then interpreted as if the actually measured state had propagated backwards in time to the preparer (see Appendix~B for more details). We consider here an intermediate picture, which we call partial time reversal, where a bipartite system is partly propagated forwards and backwards in time. The intuition behind this picture comes from comparing the Hamiltonian generating a beam-splitter transformation $H_{\rm BS}\propto \hat a_1^\dagger \hat a_2 + \hat a_1 \hat a_2^\dagger $  and a two-mode squeezer $H_{\rm TS}\propto \hat a_1^\dagger \hat a_2^\dagger + \hat a_1 \hat a_2 $, where $\hat a_1$ and $\hat a_2$ denote bosonic mode operators. Evidently, by interchanging $\hat a_2$ and $\hat a_2^\dagger$ we convert $H_{\rm BS}$ into $H_{\rm TS}$, suggesting that these two Gaussian transformations are dual under partial time reversal~\cite{cewqo}.

\begin{figure}[t]
	\includegraphics[width=3.cm]{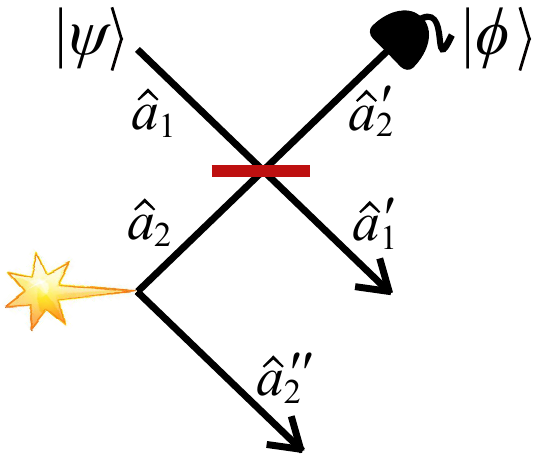}
	\caption {A beam splitter (red segment) of transmissivity $t$ is converted into a two-mode squeezer of gain $1/t$ under partial time reversal. The evolution of the first mode is described in the predictive picture ($\hat a_1 \to \hat a_1'$), while that of the second mode is expressed in the retrodictive picture ($\hat a_2' \to \hat a_2''$). The output $\hat a_2''$ is accessed via an EPR state (yellow star).
	} {\label{fig0}}
\end{figure}

More precisely, consider a beam splitter which effects the linear coupling between $\hat a_1$ and $\hat a_2$ as shown in Fig.~\ref{fig0}.  Mode $\hat a_1$ is prepared in state $\ket{\psi}$ in the predictive picture, while mode $\hat a_2$ is prepared in state $\ket{\phi}$ in the retrodictive picture (physically, the output mode $\hat a_2'$ is post-selected in state $\ket{\phi}$). A photon-number measurement on mode $\hat a_2$ in the retrodictive picture based on the resolution of identity $\sum_n \proj{n} = \openone$ would induce an (unnormalized) uniform mixture of Fock states $\ket{n}$ in mode $\hat a_2$. Physically, we need to prepare an (unnormalized) EPR state  $ \sum_n \ket{n}_{a_2} \ket{n}_{a_2''}$ and use its other leg $\hat a_2''$ as the output in the retrodictive picture, see Fig.~\ref{fig0}.
As a simple example, take $\ket{\psi}=\ket{\phi}=\ket{0}$. If modes $\hat a_2$ and $\hat a_2''$ are in state $\ket{n}$, then the joint output state is
\begin{equation}
\sum_{i=0}^n {n \choose i}^{1/2}  t^{i/2} \, (1-t)^{(n-i)/2} \, \ket{n-i}_{a_1'}  \ket{i}_{a_2'}  \ket{n}_{a_2''},
\end{equation}
with $t$ being the beam-splitter transmissivity. By post-selecting mode $\hat a_2'$ in the retrodicted state $\ket{0}$, we get the output state $\propto (1-t)^{n/2} \, \ket{n}_{a_1'}  \ket{n}_{a_2''}$. Summing over $n$, we recognize a two-mode squeezed vacuum state, $\propto \sum_{n=0}^\infty  \xi^n \, \ket{n}_{a_1'}  \ket{n}_{a_2''}$ (of parameter $\xi = \sqrt{1-t}$), which is precisely the state resulting from applying a two-mode squeezer onto $\ket{\psi}\ket{\phi}=\ket{0}\ket{0}$. This extends to any inputs $\ket{\psi}$ and $\ket{\phi}$, so we conclude that a beam splitter of transmissivity $t$ is converted into a two-mode squeezer of parameter $\xi \equiv \tanh r = \sqrt{1-t}$, i.e., of gain $g \equiv \cosh^2 r = 1/t$ (see Appendix~B). 
%The conservation of the photon number in the beam splitter straightforwardly translates into the conservation of the photon number difference in the two-mode squeezer.

Remarkably, an active transformation is thus simulatable with a passive linear-optics interferometer. Note that an EPR state is used in order to access the output mode $\hat a_2''$, so we still need an active medium. However, this is a prior resource only, and no in-line nonlinearity is needed during the process itself. Physically, the EPR state must be approached with a two-mode squeezed vacuum state, and finite squeezing manifests itself as an additional filtering in Fock basis in the circuit (as  we shall see, it can be counteracted in the simulation procedure). In contrast, if we do not need to access $\hat a_2''$, we may simply prepare $\hat a_2$ in a random state and get rid of any active medium (cf. the implementation of an optical amplifier without nonlinearity \cite{josse}). In this case, the nonlinearity solely originates from the post-selection process.

\section{Time-unfolded linear optical circuit}\label{time}

\begin{figure}[ht]
	(a) \includegraphics[width=6.5cm]{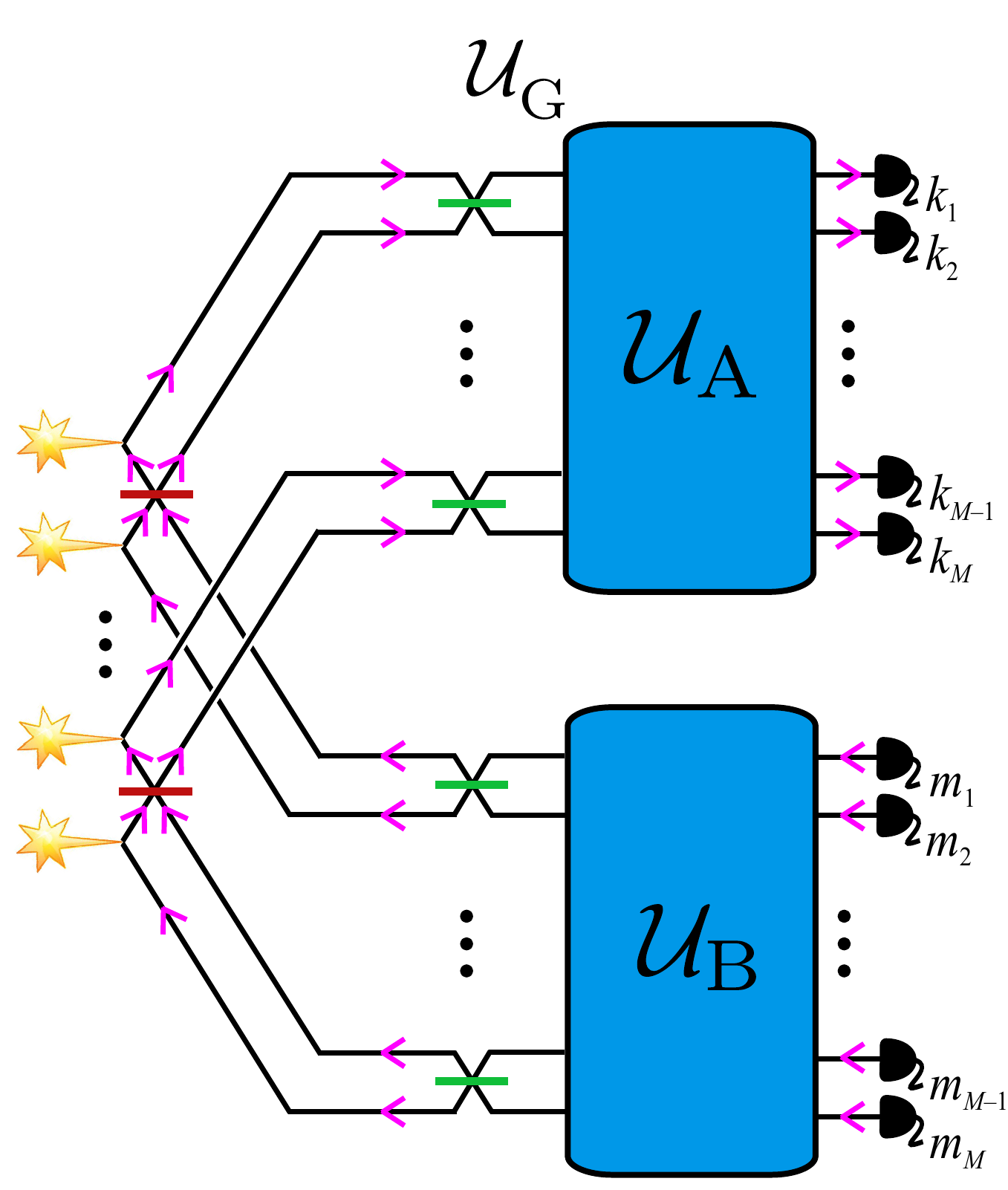}\\
	(b) \includegraphics[width=7.cm]{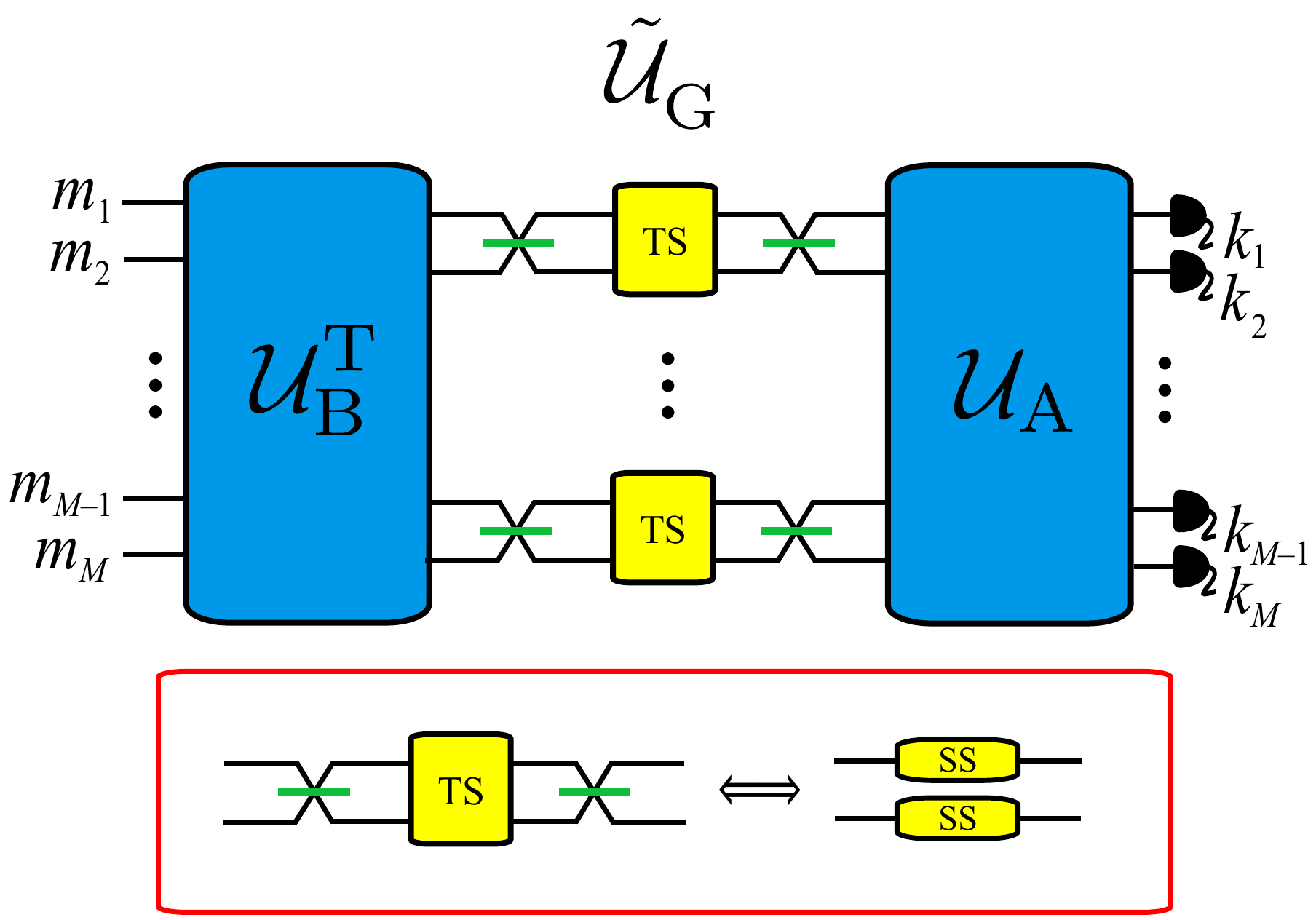}
	\caption {(a) The linear-optical circuit $\mathcal{U}_\ttt{G}$ simulates the Gaussian circuit  $\tilde{\mathcal{U}}_\ttt{G}$ depicted in panel (b). The adjacent modes of $M$ two-mode squeezed vacuum states (yellow stars) are combined pairwise on beam splitters of transmissivity $t_j$ (red segments) and then sent onto two linear-optical circuits $\mathcal{U}_\ttt{A}$ and $\mathcal{U}_\ttt{B}$, preceded by a row of balanced beam splitters (green segments). The black lines refer to the physical time evolution from left to right, while the pink arrows show the information flow in the time-unfolded picture. (b) The Gaussian circuit $\tilde{\mathcal{U}}_\ttt{G}$ corresponds to the time-unfolded version of $\mathcal{U}_\ttt{G}$. The two-mode squeezers (TS) result from partial time reversal. The inset shows the equivalence of a two-mode squeezer sandwiched between two balanced beam splitters (green segments) and two single-mode squeezers (SS).} {\label{fig1}}
\end{figure}

We now exploit partial time reversal and build a linear-optical circuit that can be mapped onto any Gaussian circuit.
Our construction utilizes a set of  $M$ equally squeezed two-mode  squeezed vacuum states (TMSs) $\ket{\psi_\ttt{in}}=\otimes_{j=1}^{M}\ket{\psi_j}$ as a resource, where each state $\ket{\psi_j}=(1-\xi^2)^{1/2} \sum_{n_j=0}^\infty \xi^{n_j}\ket{n_j}\ket{n_j}$ has a squeezing parameter $\xi$ ($0\leq \xi<1$). As illustrated in Fig.~\ref{fig1}(a), for each pair of adjacent TMSs $\{\ket{\psi_1}, \ket{\psi_2}\}, \{\ket{\psi_3}, \ket{\psi_4}\}, \dots, \{\ket{\psi_{M-1}},  \ket{\psi_{M}}\}$, we combine the lower leg of $\ket{\psi_{j}}$ with the upper leg of $\ket{\psi_{j+1}}$ on a beam splitter $\mathcal{U}_\ttt{BS}^{(j)}$ of transmissivity $t_j$ (we assume that $M$ is even). The upper legs of the emerging modes are then combined pairwise in a row of balanced beam splitters $\mathcal{U}_\ttt{BS}$ and then injected into the $M$-port linear-optical circuit $\mathcal{U}_\ttt{A}$, while  the lower legs are similarly combined pairwise in a row of balanced beam splitters and sent to $\mathcal{U}_\ttt{B}$. We call  $\mathcal{U}_\ttt{G}$ the resulting $2M$-mode linear-optical circuit, namely
\be\label{3.0}
\mathcal{U}_\ttt{G} = (\mathcal{U}_\ttt{A} \otimes  \mathcal{U}_\ttt{B}) \,\,  \mathcal{U}_\ttt{BS}^{\otimes M}
\left( \otimes_{j=1}^{M/2} \mathcal{U}_\ttt{BS}^{(j)} \right)     .
\ee

As depicted in Fig.~\ref{fig1}(a), let us consider the event of detecting a pattern of single photons ${\bf k}\equiv \{k_1, \dots, k_M\}$ at the output of circuit $\mathcal{U}_\ttt{A}$, given a set of single-photon detections ${\bf m}\equiv \{m_1, \dots, m_M\}$ at the output of $\mathcal{U}_\ttt{B}$. We will prove that this event is equivalent (in a sense that is made precise below) to the situation where the pattern ${\bf k}$ is detected at the output of an $M$-mode Gaussian circuit presented with input state $\ket{\bf m}$, as shown in Fig.~\ref{fig1}(b). The proof of this statement makes use of three building blocks: the symmetry of quantum mechanics under time reversal, the conversion of a beam-splitter transformation into a two-mode squeezer under partial time reversal and the  Bloch-Messiah reduction theorem.
%of $N_\ttt{A}$ single photons
%of $N_\ttt{B}$ single photons (upon classically processing the detection statistics as is detailed in the following section). 
%Furthermore, a slightly modified version of the linear-optical circuit $\mathcal{U}^{(1)}_\ttt{G}$ simulates {\it arbitrary} $M/2$-mode Gaussian transformations injected with single-photon input states and concluded by photon counting~\cite{footnote2}. 
%The latter allows one to decompose an arbitrary Gaussian transformation into a set of single-mode squeezers, sandwiched between two linear-optical circuits. 

We start our proof by unfolding the setup depicted in Fig.~\ref{fig1}(a). That is, we backpropagate the state $\ket{{\bf m}}$ emerging upon detection at the output of circuit $\mathcal{U}_\ttt{B}$, so that quantum information flows from the output of circuit $\mathcal{U}_\ttt{B}$ towards the output of circuit $\mathcal{U}_\ttt{A}$ [pink arrows in Fig.~\ref{fig1}(a)]. More precisely, state $\ket{{\bf m}}$ evolves through the time-reversed circuit $\mathcal{U}_\ttt{B}^\ttt{T}$ (the time reversal of the transformation $\mathcal{U}_\ttt{B}$ corresponds to its transposition in the Fock basis), followed by a set of two-mode squeezers $\mathcal{U}_\ttt{TS}^{(j)}$, sandwiched between two rows of balanced beam splitters and concluded by the circuit $\mathcal{U}_\ttt{A}$. We call the resulting $M$-mode circuit $\tilde{\mathcal{U}}_\ttt{G}$, see Fig.~\ref{fig1}(b),
\be\label{3.1}
\tilde{\mathcal{U}}_\ttt{G} = \mathcal{U}_\ttt{A}   \left[\otimes_{j=1}^{M/2}\left(\mathcal{U}_\ttt{BS}^{} \, \mathcal{U}_\ttt{TS}^{(j)} \, \mathcal{U}_\ttt{BS}^{{\rm T}}\right)\right]   \mathcal{U}_\ttt{B}^\ttt{T}  .
\ee
%That is, one of their input ports propagates backwards in time, while the other one propagates forwards in time. 
The two-mode squeezers appear in $\tilde{\mathcal{U}}_\ttt{G}$ due to the crucial fact that in the time-unfolded picture, the beam splitters $\mathcal{U}_\ttt{BS}^{(j)}$ are partially time reversed (each beam splitter of transmissivity $t_j$ is converted into a two-mode squeezer of gain $g_j=1/t_j$). We further notice that, as illustrated in the inset of Fig.~\ref{fig1}(b), a two-mode squeezer $\mathcal{U}_\ttt{TS}^{(j)}$ of gain $g_j$ preceded and followed by a balanced beam splitter is equivalent to two single-mode squeezers $\mathcal{U}_\ttt{SS}^{(j_1)}$ and $\mathcal{U}_\ttt{SS}^{(j_2)}$ of parameters $r^{(j_1)}=\ttt{arccosh}\sqrt{g_j}$ and $r^{(j_2)}=-\ttt{arccosh}\sqrt{g_j}$, namely
%\be\label{3.2}
$\mathcal{U}_\ttt{BS} \, \mathcal{U}_\ttt{TS}^{(j)} \, \mathcal{U}_\ttt{BS}^{\rm T}=\mathcal{U}_\ttt{SS}^{(j_1)}\otimes{\mathcal{U}_\ttt{SS}^{(j_2)}}$.
%\ee
Consequently, the circuit $\tilde{\mathcal{U}}_\ttt{G}$ represents an instance of the Bloch-Messiah decomposition and thus encodes a set of Gaussian transformations. Apparently, this set is restricted since $r^{(j_1)}=-r^{(j_2)}$ (in general, the decomposition requires  single-mode squeezers of arbitrary parameters). However, a slight modification of $\mathcal{U}_\ttt{G}$ allows one to achieve any $M/2$-mode Gaussian transformation, i.e., at the expense of decreasing by half the number of simulated modes. As detailed in Appendix~D, this can be achieved by replacing the $M$-mode circuit  $\mathcal{U}_\ttt{A}$ ($\mathcal{U}_\ttt{B}$) with two disjoint $M/2$-mode circuits, each of which is injected with a subset of the output modes of the row $\mathcal{U}_\ttt{BS}$, thus yielding two fully general $M/2$-mode Bloch-Messiah decompositions.

Note that in the limit $t_j=1, \forall j$, the circuit $\mathcal{U}_\ttt{G}$ reduces to twofold scattershot boson sampling~\cite{twofold}. Its time-unfolded version then simulates the {\it linear-optical} boson sampling problem. Thus, we have shown that a simple coupling of adjacent TMSs within a linear-optical circuit gives access to an extended (active) boson sampling setup. Further, the depth of the circuit $\tilde{\mathcal{U}}_\ttt{G}$ is equal to the sum of the depths of $\mathcal{U}_\ttt{A}$ and $\mathcal{U}_\ttt{B}$, which may be interesting for practical implementations (see also Ref.~\cite{DrivenBS}). 

%We finally note that the Gaussian circuit $\tilde{\mathcal{U}}_\ttt{G}^{(1)}$ does not conserve the number of photons. However, if photon detection happens immediately after the row of beam splitters $\mathcal{U}_\ttt{BS}^{(j)}$, the resulting time-unfolded circuit corresponds to a set of $M$ disjoint two-mode squeezers. For a two-mode squeezer, the photon number difference at its input is equal to the photon number difference at its output. That is, e.g., for $\tilde{U}_\ttt{G}^{(1)}$, $m_i-m_{i+1}=k_i-k_{i+1}$ ($i=1, \dots, M$).

\section{Simulation procedure}
Having demonstrated the link between our time-unfolded linear optical circuit and arbitrary Gaussian transformations, we go on to establish the procedure for simulating the photon-counting probability distribution of circuit $\tilde{\mathcal{U}}_\ttt{G}$ via that of circuit  $\mathcal{U}_\ttt{G}$. We first express the joint probability $p({\bf k},{\bf m})$ of detecting the pattern ${\bf k}$ of $N_\ttt{A}$ single photons at the output of circuit $\mathcal{U}_\ttt{A}$ and the pattern ${\bf m}$ of $N_\ttt{B}$ single photons at the output of circuit $\mathcal{U}_\ttt{B}$ (see Appendix~C),
\be\label{4.1}\nonumber
p({\bf k},{\bf m})=\left|\bra{{\bf k}} \avg{{\bf m}| \mathcal{U}_\ttt{G} |\psi_\ttt{in}}\right|^2=\frac{(1-\xi^2)^M \xi^{2N}}{\prod_{j=1}^{M/2}{t_j}} \,\tilde{p}({\bf k}|{\bf m}),
\ee
where $N_\ttt{A} = \sum_{i=1}^M k_i$, $N_\ttt{B} = \sum_{i=1}^M m_i$, $2N = N_\ttt{A}+N_\ttt{B}$, and 
$\tilde{p}({\bf k}|{\bf m})\equiv \left|\bra{\bf k}\tilde{\mathcal{U}}_\ttt{G}\ket{\bf m}\right|^2$ is the conditional probability of detecting pattern ${\bf k}$ upon evolving the input state $\ket{\bf m}$ through $\tilde{\mathcal{U}}_\ttt{G}$. Thus, we observe a proportionality factor $A({\bf k},{\bf m})\equiv (1-\xi^2)^M \xi^{2N}/\prod_{j=1}^{M/2}{t_j}$, which depends both on {\bf k} (via $N_\ttt{A}$) and {\bf m} (via $N_\ttt{B}$) and may be interpreted as an extra layer effecting filtering in the Fock basis preceeding and following the row of two-mode squeezers $\mathcal{U}_\ttt{TS}^{(j)}$ in $\tilde{\mathcal{U}}_\ttt{G}$. We compensate this factor $A({\bf k},{\bf m})$ in the simulation by adapting a random-walk sampling algorithm upon the space of single-photon detection events $\{{\bf k}, {\bf m}\}$ registered at the output of $\mathcal{U}_\ttt{G}$. More specifically, we apply the Metropolised independence sampling algorithm~\cite{MIS} (see also Note~\cite{footnote3}), which allows us to get a sampling statistics of $\{{\bf k}, {\bf m}\}$ according to $\tilde{p}({\bf k},{\bf m})=\tilde{p}_0({\bf m})\, \tilde{p}({\bf k}|{\bf m})$ starting from the actual distribution $p({\bf k},{\bf m})$. Here, $\tilde{p}_0({\bf m})$ is some arbitrary distribution of input states $\ket{{\bf m}}$ which we wish to engineer. Thus, we use $p({\bf k},{\bf m})$ as a trial distribution and $\tilde{p}({\bf k},{\bf m})=\tilde{p}_0({\bf m}) \, p({\bf k},{\bf m})/A({\bf k},{\bf m})$ as the target one. Starting from a sample $\{{\bf k}, {\bf m}\}$ obtained at the output of circuit $\mathcal{U}_\ttt{G}$, we accept the next sample $\{{\bf k'}, {\bf m'}\}$ with a transition probability
\ben\label{4.2}\nonumber
T(\{{\bf k'}, {\bf m'}\}&&|\{{\bf k}, {\bf m}\})={\rm min}\left\{1, \frac{\tilde{p}({\bf k'},{\bf m'})}{\tilde{p}({\bf k},{\bf m})} \frac{p({\bf k},{\bf m})}{p({\bf k'},{\bf m'})}\right\}\\
&&={\rm min}\left\{1, \xi^{\Delta_\ttt{A}+\Delta_B }\frac{\tilde{p}_0({\bf m'})}{\tilde{p}_0({\bf m})}\right\},
\een
where  $\Delta_\ttt{A}=\sum_{i=1}^M (k_i-k_i')$ and $\Delta_\ttt{B}=\sum_{i=1}^M (m_i-m_i')$. This procedure generates a Markov chain, which, after convergence, samples the target distribution $\tilde{p}({\bf k},{\bf m})$, hence simulates the evolution through $\tilde{\mathcal{U}}_\ttt{G}$ of an input state $\ket{{\bf m}}$ taken at random from $\tilde{p}_0({\bf m})$. Remark that the outlined post-processing algorithm can also be considered as generalized post-selection: post-selection involves discarding results that do not match the corresponding conditions, while we deal with an acceptance-based random-walk on the space of detected events.
	%We also note that the average acceptance rate of the random-walk sampling algorithm is necessarily larger than 1/2 , which favours its rapid convergence.}

The probability distribution $\tilde{p}_0({\bf m})$ can be chosen beforehand. For instance, we may consider a setup where input states $\ket{{\bf m}}$ are uniformly distributed over a shell with a fixed photon number $\cal N$ by setting $\tilde{p}_0({\bf m})=\delta_{N_\ttt{B},\, \cal N}/\binom{{\cal N}+M-1}{\cal N}$. In this case, $T(\{{\bf k'}, {\bf m'}\}|\{{\bf k}, {\bf m}\})={\rm min}\{1, \xi^{\Delta_\ttt{A}}\}$. Such a scenario is in the spirit of scattershot boson sampling~\cite{scattershot}. In particular, in the limit $t_j=1, \forall j$, our setting is equivalent to twofold scattershot boson sampling~\cite{twofold} and $\Delta_\ttt{A}=\Delta_{\ttt{B}}=0$. Hence, every sample from the circuit $\mathcal{U}_\ttt{G}$ is accepted. We also emphasize that choosing $\tilde{p}_0({\bf m})$ as a Gibbs distribution is equivalent to simulating an ensemble of thermally excited bosonic modes. Given the analogy between photons distributed among optical modes and molecular phonons among vibrational modes, this approach is highly relevant to simulating spectra of molecular vibronic transitions at a non-zero temperature \cite{molec}.  

\section{Conclusion}

In this paper, we report on a linear-optical scheme for simulating sampling from arbitrary Gaussian circuits. We showed that by making use of two-mode squeezed vacuum states as a prior resource, such a simulation of the Bloch-Messiah decomposition of an arbitrary Gaussian transformation can be achieved with linear optics. Our setup therefore shares some similarities with (a Gaussian counterpart of) the KLM scheme:
%; the latter simulates arbitrary quantum circuits with linear optics supplemented with single-photon sources and detection. 
here, we simulate (at a sampling level) Gaussian circuits with linear optics (with the additional need for prior Gaussian entanglement but no need for ancillary resources). The building block of our construction is the equivalence between a two-mode squeezer and a partially time-reversed beam splitter. Time-symmetry considerations also play a main role in our construction, demonstrating once again how the notion of time reversal can contribute to the development of quantum computing~\cite{twofold}. We also introduce a post-processing random-walk sampling algorithm, which can be considered as generalized post-selection. This probabilistic algorithm has an average acceptance rate larger than 1/2 independently of the number of photons and modes involved  for a certain set of probability distributions, which ensures fair convergence (see also Appendix~C).

Our work identifies a new class of quantum circuitry with post-processing that yields a specialized -- Gaussian -- set of programmable simulators. This result contributes to the understanding of the hierarchy of restricted photonic non-linearities. In fact, our sub-universal scheme lies in-between purely linear optical and full-fledged universal photonic setups. We therefore expect that it may be better suited to achieve the regime of quantum advantage (despite the need for data post-processing) and may find more practical applications than, e.g., the original boson sampling setup. In particular, we believe that our scheme can be utilized for quantum simulations of molecular spectra, deep neural networks and in quantum metrology, where single-mode non-linearities and Gaussian operations play a crucial role~\cite{molec, neural, metrology}. For instance, given the analogy between photonic and phononic modes, our setting makes a natural platform for the implementation of Duschinsky rotations, which consist of two passive transformations intermitted by single-mode squeezers~\cite{molec, doktorov}. In addition, our post-processing random-walk sampling algorithm allows one to engineer arbitrary prior (e.g., thermal) distribution of input vibrational excitations. Hence, the proposed circuit can be seen as a tool for molecular vibronic spectra simulations at non-zero temperatures. From the resource point of view, simulating a given number of vibrational modes would require twice as many two-mode squeezed vacuum states (current photonic simulations deal with up to six-mode molecules~\cite{molec_exp, exp_anthony}, while computationally hard simulations are expected to require more than ten modes~\cite{class, class1}).

Importantly, since our simulation scheme involves non-Gaussian resources in the form of Fock states and photon counting, it goes beyond the classically simulatable Gaussian computational model~\cite{knill}. (Note that even classically simulatable but non-trivial setups can be beneficial to the development of quantum-inspired classical computational algorithms~\cite{algo}.) Furthermore, our scheme might be generalizable to a setting with arbitrary input states and detection, including a hybrid combination of discrete- and continuous-variable resources, which may yield another path for generalizing the boson sampling paradigm. It also finds additional connection to measurement-based continuous-variable quantum computing, where prior Gaussian resources can be used for building cluster states~\cite{MBQC}.
%We plan to address this question, as well as the computational complexity of the corresponding problems in a future work.

Finally, we emphasize that current and emerging integrated photonic technologies, such as Lithium-Niobate and silicon based photonic hardware, are candidate platforms for the realization of our scheme. The on-chip strong non-linearities enable one to generate resource two-mode squeezed vacuum states via non-degenerate three-wave mixing (in Lithium-Niobate) or degenerate spontaneous four-wave mixing (in silicon) processes. The state evolution and detection can, in turn, be implemented by means of  manufacturable programmable photonic circuitry and (integrated) superconducting photo-detectors~\cite{LiN, detect, integrated}. We remark, however, that optical losses remain a crucial challenge in this context. In particular, it is important to realize how the circuit transmission and coupling losses (e.g., chip-to-fiber coupling loss for off-chip detection), as well as detection inefficiencies will affect experimental fidelities (unlike finite squeezing, losses cannot be compensated in a straightforward way via our data post-processing algorithm).  Nevertheless, we believe that current silicon and Lithium-Niobate based (reconfigurable) integrated photonic technologies with $\approx 0.2$dB loss per beam splitter transformations, robust multi-photon interferometers reaching 99$\%$ efficiencies for circuits with up to tens of modes~\cite{janpan, stefano, LiN, jwpan}, low-loss grating couplers~\cite{grating} and superconducting nano-wire single-photon detectors (reaching 70-80\% efficiencies) suggest the feasibility of our proposed architecture, at least at a moderate-size level. We will further address the practicability of our setting with existing photonic platforms in a future work.

\section*{Acknowledgments}
The authors thank Anthony Laing, Stefano Paesani, Timothy C. Ralph, Raffaele Santagati and Jianwei Wang for useful discussions and comments. This work was supported by the F.R.S.-FNRS Foundation under Project No. T.0224.18.

%This work was supported by the H2020-FETPROACT-2014 Grant QUCHIP (Quantum Simulation on a Photonic Chip, grant agreement no. 641039) and by the F.R.S.-FNRS Foundation under Project No. PDR T.0224.18. 

%as we shall demonstrate via the retrodictive approach to quantum mechanics~\cite{retro}.

%Moreover, certain nonlinear operations outside of the Gaussian domain can be approximated to a high degree of accuracy by Gaussian transformations \LC{[reference?]}. 

\appendix

\section{Phase-space representation}\label{phase}

We here recall the phase-space description of Gaussian states and transformations. Any $M$-mode Gaussian state can be described in terms of its $2M\times 2M$ covariance matrix $\sigma_{\ttt{in}}$ with matrix elements defined as
\be
\sigma_{ij}^{(\ttt{in})}=\frac{1}{2}\avg{\{\hat{R}_i,\hat{R}_j^\dagger\}}-\avg{\hat{R}_i}\avg{\hat{R}_j}^*,
\ee
where the $2M$-component vector $\hat{R}\equiv\{\hat{a}_1, \dots, \hat{a}_M, \hat{a}_1^\dagger, \dots, \hat{a}_M^\dagger\}$ contains the $M$ creation and annihilation operators of the photonic modes (we are interested here in states with zero displacement only, i.e., $\avg{R_i}=0, \forall i$). The Gaussian evolution of a state may be expressed in terms of the evolution of its covariance matrix $\sigma$, namely
\be
\sigma_{\ttt{out}}=S\sigma_{\ttt{in}}S^\dagger,
\ee
where $S$ is the complex symplectic matrix that defines the Gaussian transformation and satisfies $S\Sigma S^\dagger=\Sigma$, with $\Sigma=\begin{bmatrix}
I_M & 0  \\
0 & -I_M 
\end{bmatrix}$ ($I_M$ is the $M\times M$ identity matrix). The matrix $S$ can also be seen as the transformation that maps the input mode operators $\hat{R}_l$ (with $l=1,\dots, 2M$) onto the output mode operators $\hat{Q}_k$ (with $k=1,\dots, 2M$): 
\begin{equation}\label{1}
\hat{Q}_k=\sum_{l=1}^{2M}  S_{{kl}} \, \hat{R}_l.
\end{equation}
The symplectic matrix $S$ therefore encapsulates the phase-space representation of the corresponding Gaussian transformation. 

In this work, we are interested in three types of Gaussian transformations: beam-splitter transformation, two-mode and single-mode squeezers. In phase-space representation, a beam-splitter transformation of transmissivity $t_j$ is defined as
\ben
&& S_\ttt{BS}^{(j)}=\begin{bmatrix}
U_\ttt{BS}^{(j)} & 0  \\
	0 & U_\ttt{BS}^{{(j)}}
\end{bmatrix},\\
&& U_\ttt{BS}^{(j)}=\begin{bmatrix}
	\sqrt{t_j} & \sqrt{1-t_j}  \\
	-\sqrt{1-t_j} & \sqrt{t_j} 
\end{bmatrix}.
\een
The beam-splitter is a linear-optical (or passive) transformation, which means that it conserves the number of photons. Note also that an arbitrary $M$-mode linear-optical (passive) transformation can be decomposed into a set of beam splitters and phase shifters~\cite{reck}.

The phase-space representation of a two-mode squeezer of gain $g_j$ reads
\ben
&& S_\ttt{TS}^{(j)}=\begin{bmatrix}
	\sqrt{g_j} & 0  & 0 & \sqrt{g_j-1}\\
	0 & \sqrt{g_j} & \sqrt{g_j-1}& 0 \\
	0 & \sqrt{g_j-1}  & \sqrt{g_j} & 0\\
	\sqrt{g_j-1} & 0 & 0& \sqrt{g_j}
\end{bmatrix}.
\een
The gain is related to the squeezing parameter $\xi_j$ via the relation $g_j=(1- \xi_j^2)^{-1}$. 
Two-mode squeezing operation does not conserve the total photon number, but it does conserve the difference of the input and output photon numbers. That is, if a two-mode squeezer is injected with the Fock state $\ket{m_1, m_2}$ and the state $\ket{k_1,k_2}$ is detected at its output, then $m_1-m_2=k_1-k_2$.

Finally, the single-mode squeezer of a squeezing degree $r^{(j)}$ has the following phase-space representation
\ben
&& S_\ttt{SS}^{(j)}=\begin{bmatrix}
    \ttt{cosh}\,{r^{(j)}} & \ttt{sinh}\, {r^{(j)}}  \\
	\ttt{sinh}\,{r^{(j)}} & \ttt{cosh}\,{r^{(j)}}
\end{bmatrix}.
\een
Importantly, due to the Bloch-Messiah reduction theorem~\cite{BM}, an arbitrary $M$-mode Gaussian transformation $S$ can be represented as a set of single-mode squeezers, sandwiched between two $M$-mode linear-optical (passive) circuits $S_1$ and $S_2$:
\ben
S=S_1 \left[\oplus_{j=1}^MS_\ttt{SS}^{(j)}\right]S_2.
\een

\section{Beam splitter under partial time reversal}\label{beam}

In order to make the notion of partial time reversal more precise, we must first recall the retrodictive picture of quantum mechanics \cite{aharonov1964}, which is the time-reversed version of the (usual) predictive picture. In the latter picture, one makes predictions about the outcomes of some POVM measurement $\{ \Pi_n\}$ from the prior knowledge of the state $\rho_m$ (prepared with probability $p_m$). Born's rule then gives us the conditional probabilities $P(n|m)= \mathrm{Tr}(\rho_m \Pi_n)$. In the retrodictive picture, one takes the opposite viewpoint and starts from the actually observed outcome $n$ (which is associated with a retrodicted state $\sigma_n$), and makes retrodictions about the preparation of the system by applying a POVM measurement $\{ \Theta_m \}$ (whose outcome $m$ is associated with the prepared state $\rho_m$). Choosing  $\sigma_n = \Pi_n /  \mathrm{Tr} (\Pi_n)$ and $\Theta_m  \propto p_m \rho_m$ (assuming that $\sum_m p_m \rho_m \propto \openone$), we may apply Born's rule in the backwards direction and get conditional probabilities  $P(m|n) = \mathrm{Tr}(\sigma_n \Theta_m)$ which are consistent with Bayes rule.

The retrodictive picture can be successfully exploited in quantum optics, see, e.g., Ref. \cite{S-retro}. As a simple illustration, let us consider the preparation of a coherent state $\rho_\alpha= \proj{\alpha}$ followed by its photon-number measurement, associated with $\Pi_n=\proj{n}$. The conditional probability of observing $n$ when preparing $\alpha$ is 
\begin{equation}
p(n|\alpha) = \mathrm{Tr}(\rho_\alpha \Pi_n) = \frac {e^{-|\alpha|^2} |\alpha|^{2n} } {n!}
\label{equS1}
\end{equation}
with $\sum_n p(n|\alpha) = 1$, $\forall \alpha$. In the retrodictive picture, one prepares the retrodicted state $\sigma_n = \proj{n}$ and applies the (continuous) POVM measurement  $\Theta_\alpha =  p_\alpha  \proj{\alpha}$, using the resolution of identity $\int {\rm d}^2\alpha \, p_\alpha \, \rho_\alpha = \openone$ with the (unnormalized) probability $p_\alpha=1/\pi$. Thus, we backpropagate a number state and apply an eight-port homodyne (also called heterodyne) detection, resulting in the conditional probabilities
\begin{equation}
p(\alpha|n) = \mathrm{Tr}(\sigma_n \Theta_\alpha) =  \frac {e^{-|\alpha|^2} |\alpha|^{2n} } {\pi \, n!}
\label{equS2}
\end{equation}
with $\int {\rm d}^2\alpha \, p(\alpha|n) = 1$, $\forall n$.
The connection between the probability distribution (\ref{equS1}) of measuring a given photon number $n$ in a coherent state and the probability density (\ref{equS2}) of measuring a specific $\alpha$ in a Fock state originates from the duality between the predictive and retrodictive pictures.

\begin{figure}[t]
	\includegraphics[width=3cm]{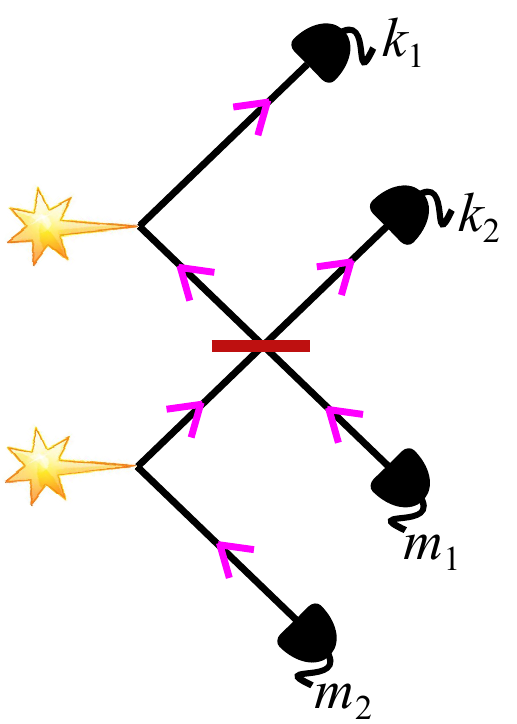}
	\caption {A beam splitter of transmissivity $t$ (red segment) is fed by two modes originating each from a two-mode squeezed vacuum state (yellow stars), and we focus on the probability of the photon-counting event $k_1,k_2,m_1,m_2$. By unfolding this linear optical circuit in time, the two-mode squeezed vacuum states can be viewed as ``wires'', and we get a two-mode squeezer of gain $1/t$ with inputs $m_1,m_2$ and outputs $k_1,k_2$ (the pink arrows indicate the information flow in this time-unfolded picture). In order to account for the finite squeezing of the two-mode squeezed vacuum states, the two-mode squeezer of gain $1/t$ must actually be preceded and followed by a filtering operation in Fock basis. } {\label{figS0}}
\end{figure}

Coming back to partial time reversal, we now investigate an intermediate situation involving two bosonic modes, one of them being described in the predictive picture while the other is analyzed in the retrodictive picture. In the main text, we have seen that a beam splitter of transmissivity $t$ is converted into a two-mode squeezer with gain $g=1/t$ under such a partial time-reversal~\cite{cewqo}. Let us now show that, by adding a second EPR state on the other input mode, we reach a symmetric scheme that can be used as a building block in our simulation procedure, see Fig.~\ref{figS0}. 
%\LC{Note that while in Fig.~1 of the main text one of the inputs of the beam splitter is a “conventional” input and the other one is a “time-reversed” input implemented via a measurement, in Fig.~\ref{figS0} both inputs are “time-reversed” (i.e., implemented via measurement). Consequently, they need to be of the same kind, which necessitates an extra two-mode squeezed vacuum state as compared to the situation of Fig.~1. This explanation applies, as well, to Fig.~2(a), of the main text, where the circuit $\mathcal{U}_\ttt{G}$ requires $M$ two-mode squeezed vacuum input states.} 
	
Consider now two 2-mode squeezed vacuum states
\ben
\ket{\psi_\ttt{in}} = (1-\xi^2) \sum_{n_1,n_2=0}^\infty \xi^{n_1+n_2} \ket{n_1,n_1,n_2,n_2} 
\een
and let us express the probability amplitude corresponding to the detection pattern $\{k_1,k_2,m_1,m_2\}$ shown in Fig.~\ref{figS0}, namely
\ben
\bra{k_1,k_2,m_1,m_2}&&  \mathcal{U}_\ttt{BS}^{(t)}  \ket{\psi_\ttt{in}}  =  \\ \nonumber
&&(1-\xi^2) \xi^{k_1+m_2} \bra{m_1,k_2}  \mathcal{U}_\ttt{BS}^{(t)} \ket{k_1,m_2}
\een
where  $\mathcal{U}_\ttt{BS}^{(t)}$ is a beam splitter of transmissivity $t$. Using the correspondence with a two-mode squeezer $\mathcal{U}_\ttt{TS}^{(1/t)}$ of gain $1/t$, namely,
\ben
\bra{m_1,k_2}  \mathcal{U}_\ttt{BS}^{(t)} \ket{k_1,m_2} = \frac {1}{\sqrt{t}}   \bra{k_1,k_2}  \mathcal{U}_\ttt{TS}^{(1/t)} \ket{m_1,m_2},\,\,\,\,
\een
we can reverse time for the first mode ($k_1 \leftrightarrow m_1$). Thus, the probability of detecting pattern $\{k_1,k_2,m_1,m_2\}$ can be written as
\ben
p(k_1,k_2, && m_1,m_2)  =  \\ \nonumber
&&\frac {(1-\xi^2)^2 \xi^{2(k_1+m_2)} } {t}  \left|\bra{k_1,k_2}  \mathcal{U}_\ttt{TS}^{(1/t)} \ket{m_1,m_2}\right|^2.
\een
By photon number conservation in the beam splitter, we have
\ben
k_1+m_2 = k_2+m_1 = (k_1+k_2+m_1+m_2)/2,
\een
so that the probability of detecting pattern $\{k_1,k_2,m_1,m_2\}$ becomes
\ben
p(k_1,k_2,&& m_1,m_2)  =  \\ \nonumber
&&\frac {(1-\xi^2)^2 } {t}  \bra{k_1,k_2} \xi^{\hat n_1+\hat n_2} \, \mathcal{U}_\ttt{TS}^{(1/t)} \, \xi^{\hat n_1+\hat n_2} \ket{m_1,m_2}.
\een
It is thus proportional to the probability of detecting the pattern  $\{k_1,k_2\}$ at the output of the two-mode squeezer $\mathcal{U}_\ttt{TS}^{(1/t)} $ when sending 
the input pattern $\{m_1,m_2\}$, except for the fact that a filtration operation in Fock basis $\xi^{\hat n_1+\hat n_2}$ must be inserted before and after the two-mode squeezer. This filtration accounts for the finite squeezing of the input two-mode squeezed vacuum states, and must be compensated in the simulation algorithm as explained in the main text. 

\begin{figure}[ht]
	(a) \includegraphics[width=6.5cm]{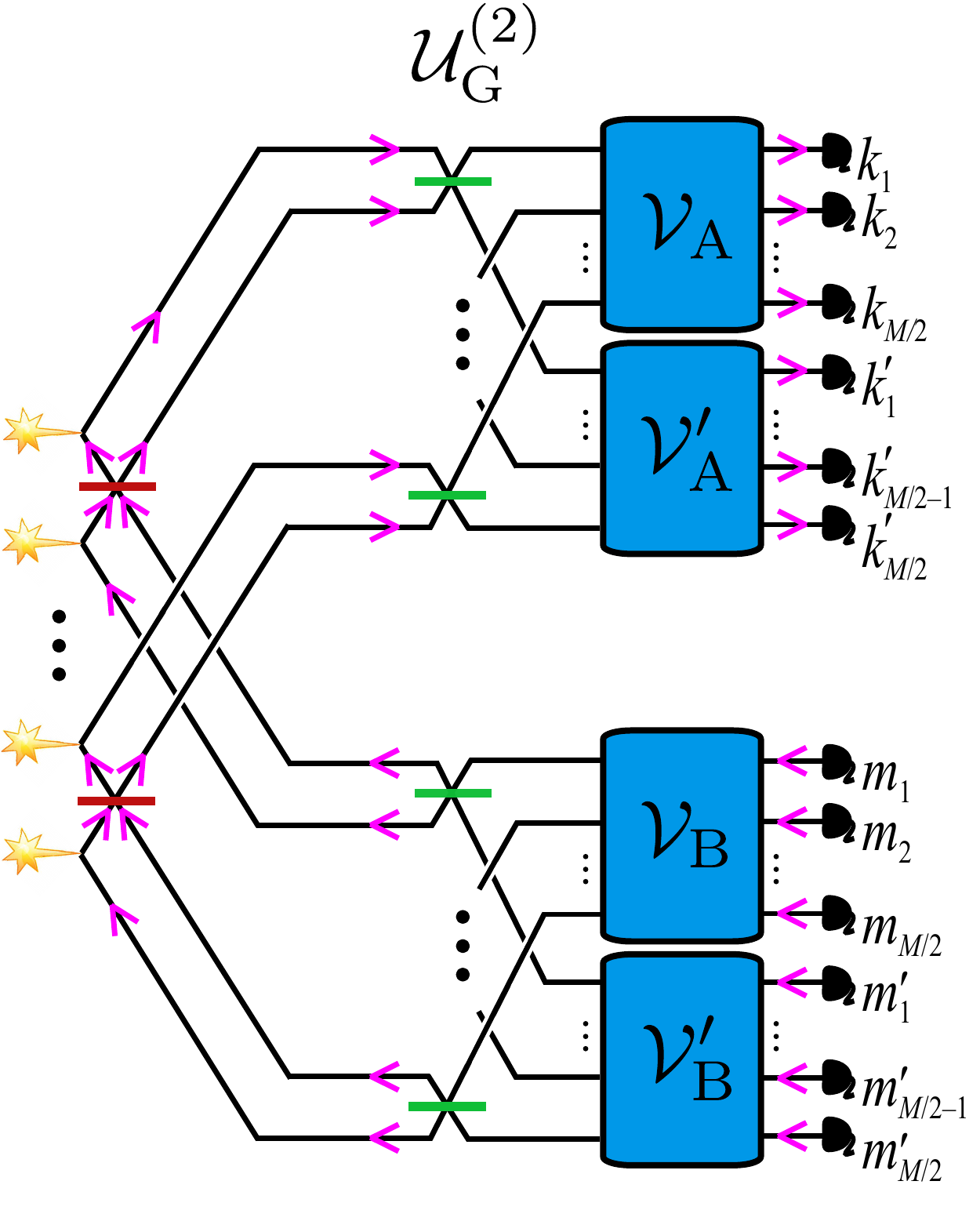}\\
	(b) \includegraphics[width=6.5cm]{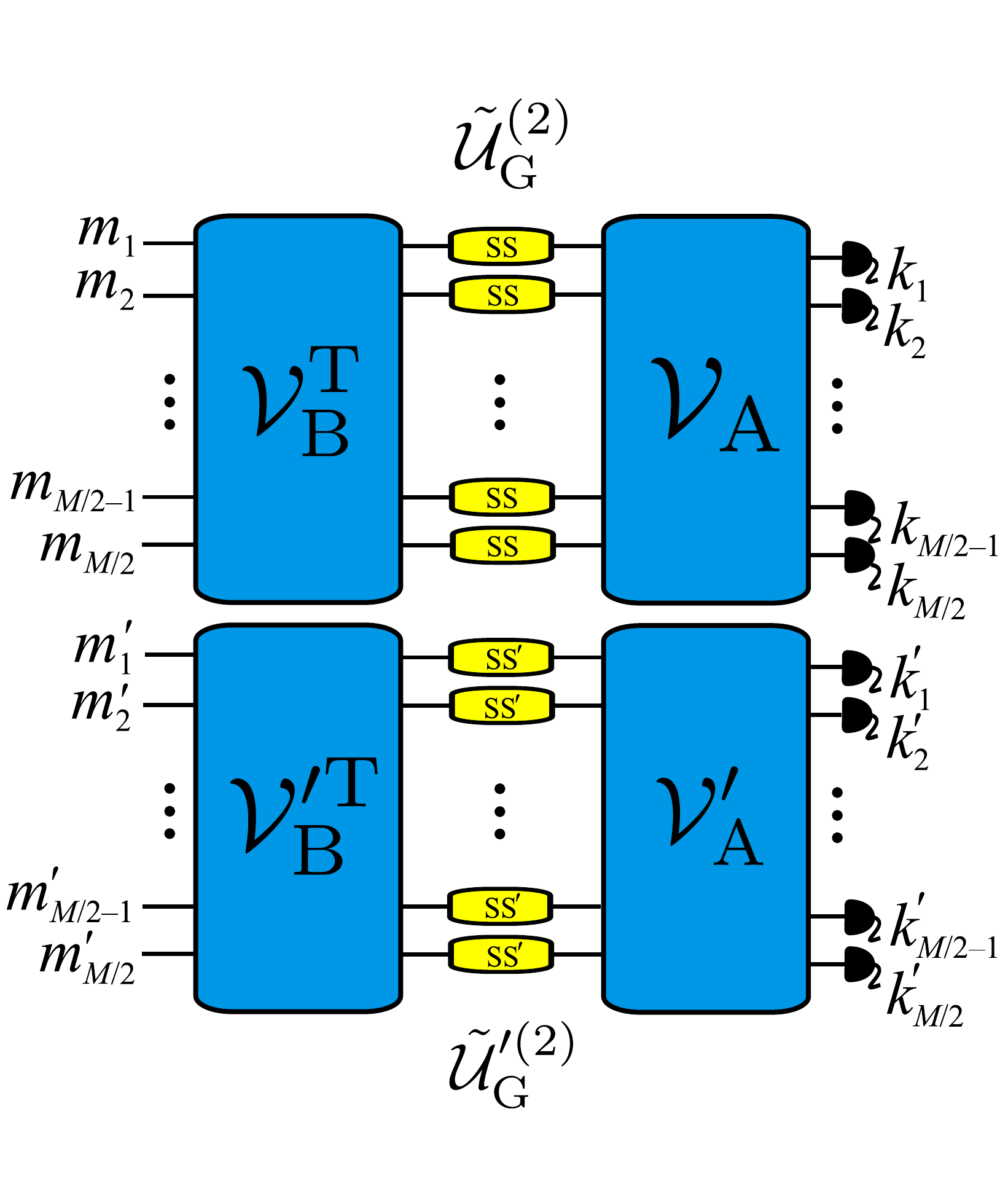}
	\caption {(a) The depicted linear-optical circuit $\mathcal{U}_\ttt{G}^{(2)}$, which is a slight modification of the circuit $\mathcal{U}_\ttt{G}$ [cf. Fig.~2(a)], simulates arbitrary Gaussian transformations. Here, the lines and arrows have the same meaning as in Fig.~2(a): the black lines illustrate the evolution of input TMSs, which propagate from left to right, the pink arrows show the information flow in our time-unfolded formalism; (b) The time unfolded version of the circuit $\mathcal{U}_\ttt{G}^{(2)}$ is equivalent to two disjoint {\it arbitrary} Gaussian transformations $\tilde{\mathcal{U}}_\ttt{G}^{(2)}$ and $\tilde{\mathcal{U'}}_\ttt{G}^{(2)}$, each achieved by means of the Bloch-Messiah decomposition.} {\label{figS1}}
\end{figure}

Remark that while in Fig.~1 one of the two modes feeding the beam splitter is a “conventional” input (state $\ket{\psi}$) and the other one is a “time-reversed” input (state $\ket{\phi}$) implemented via a measurement, in Fig.~2 (as well as in Fig.~\ref{figS1}) both inputs of the equivalent two-mode squeezer are “time-reversed” (i.e., implemented via a measurement). This is because we need both these input modes of the equivalent two-mode squeezer to emerge from a unitary $U_\ttt{B}^\ttt{T}$ in the (time-unfolded) circuit of Fig.~2(b). Consequently, they need to be both of the same kind in Fig.~2(a) (and in the building-block depicted in Fig.~\ref{figS0}), i.e., both are “time-reversed”, which thus necessitates an extra two-mode squeezed vacuum state as compared to the situation of Fig. 1.

\section{Photon-counting probability distribution of the time-unfolded linear-optical circuit}\label{proof}

We here prove the relation between photon-counting probability distributions of circuits $\mathcal{U}_\ttt{G}$ and $\tilde{\mathcal{U}}_\ttt{G}$ defined in the main text and depicted in Figs.~2(a) and~(b). Consider the joint probability $p({\bf k},{\bf m})$ of detecting the pattern of single photons ${\bf k}\equiv \{k_1, \dots, k_M\}$ at the output of the circuit $\mathcal{U}_\ttt{A}$ and the pattern ${\bf m}\equiv \{m_1, \dots, m_M\}$ at the output of $\mathcal{U}_\ttt{B}$ ($\sum_{i=1}^M k_i\equiv N_\ttt{A}$, $\sum_{i=1}^M m_i\equiv N_\ttt{B}$, and $N_\ttt{A}+N_\ttt{B}\equiv 2N$),

\begin{widetext}
\ben\label{app1} \nonumber
p({\bf k},{\bf m})= && \left|\bra{{\bf k}} \avg{{\bf m}|\mathcal{U}_\ttt{G}|\psi_\ttt{in}}\right|^2=\prod_{i=1}^M(1-\xi_i^2) \left|\sum_{{n_1, \dots, n_M}=0}^{\infty}\xi_1^{n_1}\cdots \xi_M^{n_M} \bra{\bf k}\bra{\bf m}\left[\mathcal{W}_\ttt{A}\otimes \mathcal{W}_\ttt{B}\right]\left[\mathcal{U}_\ttt{BS}^{(1)}\otimes \cdots \otimes \mathcal{U}_\ttt{BS}^{(M/2)}\right]\ket{\bf n}\ket{\bf n}\right|^2=\\ 
&&(1-\xi^2)^M \xi^{2N} \left|\sum_{n_1, \dots, n_M=0 }^{\infty}\sum_{\substack{p_1, \dots, p_M=0\\ q_1, \dots, q_M=0}}^{\infty}\bra{\bf k}\mathcal{W}_\ttt{A}\ket{\bf q}\bra{\bf m}\mathcal{W}_\ttt{B}\ket{\bf p}\bra{\bf p}\bra{\bf q}\left[\mathcal{U}_\ttt{BS}^{(1)}\otimes \cdots \otimes \mathcal{U}_\ttt{BS}^{(M/2)}\right]\ket{\bf n}\ket{\bf n}\right|^2.
\een
\end{widetext}
where $\mathcal{W}_\ttt{A}=\mathcal{U}_\ttt{A} \mathcal{U}_\ttt{BS}^{\otimes M/2}$ and  $\mathcal{W}_\ttt{B}=\mathcal{U}_\ttt{B} \mathcal{U}_\ttt{BS}^{\otimes M/2}$ $\left(\mathcal{U}_\ttt{BS}^{\otimes M/2}\right.$ is the transformation corresponding to the row of balanced beam-splitters preceding the linear-optical circuits $\mathcal{U}_\ttt{A}$ and $\mathcal{U}_\ttt{B}$). In turn, $\mathcal{U}_\ttt{BS}^{(j)}$ and $\mathcal{U}_\ttt{BS}$ stand for the beam-splitter transformation of transmissivity $t_j$ and the balanced beam-splitter transformation, respectively [cf. Fig.~2(a) and the main text]. In the above equation we have also assumed that the squeezing degrees of all two-mode squeezed vacuum states (TMSs) are equal, i.e., $\xi_1=\dots=\xi_M\equiv \xi$. Further, due to the linearity of the circuit $\mathcal{U}_\ttt{G}$, it conserves the total photon number: $2\sum_{i=1}^M n_i=\sum_{i=1}^M m_i+\sum_{i=1}^M k_i=N_\ttt{A}+N_\ttt{B}\equiv 2N$. Finally, we  have also introduced the closure relation for Fock states $\ket{{\bf p}}\equiv \ket{p_1, \dots, p_M}$ and $\ket{{\bf q}}\equiv \ket{q_1, \dots, q_M}$ in Eq.~(\ref{app1}). Remark that each $\mathcal{U}_\ttt{BS}^{(j)}$ acts on the Hilbert space of the lower leg of the $j$th TMS and the upper leg of the $(j+1)$th TMS, while $\mathcal{W}_\ttt{A}$ ($\mathcal{W}_\ttt{B}$) acts on the Hilbert space of the upper (lower) set of the emerging modes.

Now, we rewrite Eq.~(\ref{app1}) in the time-unfolded picture introduced in the main text. That is, we first take into account that the time reversal of the transformation $\mathcal{W}_\ttt{B}$ corresponds to its transposition in the Fock basis, $\bra{\bf m}\mathcal{W}_\ttt{B}\ket{\bf p}=\bra{\bf p}\mathcal{W}_\ttt{B}^\ttt{T}\ket{\bf m}$. Second, we recall that under partial time reversal a beam-splitter transformation $\mathcal{U}_{\ttt{BS}}^{(j)}$ of transmissivity $t_j$ is converted into a two-mode squeezer $\mathcal{U}_{\ttt{TS}}^{(j)}$ of a gain $g_j=1/t_j$, i.e., $\avg{c_1c_2|\mathcal{U}_{\ttt{BS}}^{(j)}|d_1d_2}=1/\sqrt{t_j}\avg{d_1c_2|\mathcal{U}_{\ttt{TS}}^{(j)}|c_1d_2}$ for any $c_1, c_2, d_1$ and $d_2$. Performing this manipulation for every beam-splitter transformation $\mathcal{U}_{\ttt{BS}}^{(j)}$ and taking into account that $\avg{n_i|p_j}=\delta_{n_i, p_j}$ and $\avg{n_i|q_j}=\delta_{n_i,q_j}$ (with $\delta_{i,j}$ being the Kronecker delta), we arrive at the following expression
\begin{widetext}
\ben\label{app2} \nonumber
p({\bf k},{\bf m})= &&
\frac{(1-\xi^2)^M \xi^{2N}}{\prod_{j=1}^{M/2}{t_j}}\left|\sum_{\substack{p_1, \dots, p_M=0\\ q_1, \dots, q_M=0}}^{\infty}
\bra{\bf k}\mathcal{W}_\ttt{A}\ket{\bf q}\bra{\bf q}\left[\mathcal{U}_\ttt{TS}^{(1)}\otimes \cdots \otimes \mathcal{U}_\ttt{TS}^{(M/2)}\right]\ket{\bf p} \bra{\bf p}\mathcal{W}_\ttt{B}^\ttt{T}\ket{\bf m}\right|^2=\\
&&\frac{(1-\xi^2)^M \xi^{2N}}{\prod_{j=1}^{M/2}{t_j}} \left|
\bra{\bf k}\mathcal{W}_\ttt{A}\left[\mathcal{U}_\ttt{TS}^{(1)}\otimes \cdots \otimes \mathcal{U}_\ttt{TS}^{(M/2)}\right]\mathcal{W}_\ttt{B}^\ttt{T}\ket{\bf m}\right|^2.
\een
\end{widetext}
Finally, we recall that a two-mode squeezer $\mathcal{U}_\ttt{TS}^{(j)}$, preceded and followed by two balanced beam splitters, is equivalent to two single-mode squeezers $\mathcal{U}_\ttt{SS}^{(j_1)}$ and $\mathcal{U}_\ttt{SS}^{(j_2)}$ of squeezing degrees $r^{(j_1)}_\ttt{S}=\ttt{arccosh}\sqrt{g_j}$ and $r^{(j_2)}_\ttt{S}=-\ttt{arccosh}\sqrt{g_j}$,
\be
\mathcal{U}_\ttt{BS}\mathcal{U}_\ttt{TS}^{(j)}\mathcal{U}_\ttt{BS}^\ttt{T}=\mathcal{U}_\ttt{SS}^{(j_1)}\otimes{\mathcal{U}_\ttt{SS}^{(j_2)}}.
\ee
Consequently, Eq.~(\ref{app2}) reads
\begin{widetext}
\ben\label{app3} \nonumber
p({\bf k},{\bf m})=\frac{(1-\xi^2)^M \xi^{2N}}{\prod_{j=1}^{M/2}{t_j}}&& \left|
\bra{\bf k}\mathcal{W}_\ttt{A}\left[ \otimes_{j_1=j_2=1}^{M/2}\left( \mathcal{U}_\ttt{SS}^{(j_1)}\otimes{\mathcal{U}}_\ttt{SS}^{(j_2)}\right)\right]\mathcal{W}_\ttt{B}^\ttt{T}\ket{\bf m}\right|^2=\\
&&\frac{(1-\xi^2)^M \xi^{2N}}{\prod_{j=1}^{M/2}{t_j}}  \left|
\bra{\bf k}\tilde{\mathcal{U}}_\ttt{G}\ket{\bf m}\right|^2\equiv A({\bf k},{\bf m}) \, \tilde{p}({\bf k}|{\bf m}).
\een
\end{widetext}
Here, $\tilde{p}({\bf k}|{\bf m})\equiv \left|\bra{\bf k}\tilde{\mathcal{U}}_\ttt{G}\ket{\bf m}\right|^2$ is the conditional probability of detecting the photon pattern ${\bf k}$ upon the Gaussian evolution $\tilde{\mathcal{U}}_\ttt{G}$ of the input state $\ket{\bf m}$. Recall that our goal is to simulate sampling from the probability distribution $\tilde{p}({\bf k},{\bf m})=\tilde{p}_0({\bf m})\tilde{p}({\bf k}|{\bf m})$, where  $\tilde{p}_0({\bf m})$ is a specific (arbitrarily chosen) probability distribution over input states $\ket{{\bf m}}$, which we are able to engineer. As explained in the main text, to achieve this we adopt a random-walk sampling algorithm, taking into account that $\tilde{p}({\bf k},{\bf m})=\tilde{p}_0({\bf m})p({\bf k},{\bf m})/A({\bf k},{\bf m})$. That is, the algorithm acts upon the space of photon detection events $\{{\bf k}, {\bf m}\}$ registered at the output of $\mathcal{U}_\ttt{G}$ and we consider $p({\bf k},{\bf m})$ as the proposal distribution while $\tilde{p}({\bf k},{\bf m})$ is the target distribution. Consequently, starting from a tuple $\{{\bf k}, {\bf m}\}$ obtained via the circuit $\mathcal{U}_\ttt{G}$, we accept the next sample $\{{\bf k'}, {\bf m'}\}$ with a transition probability
\ben\label{4.2}\nonumber
T(\{{\bf k'}, {\bf m'}\}|\{{\bf k}, {\bf m}\})= && {\rm min}\left\{1, \frac{\tilde{p}({\bf k'},{\bf m'})}{\tilde{p}({\bf k},{\bf m})} \frac{p({\bf k},{\bf m})}{p({\bf k'},{\bf m'})}\right\}=\\
&& {\rm min}\left\{1, \xi^{\Delta_\ttt{A}+\Delta_B }\frac{\tilde{p}_0({\bf m'})}{\tilde{p}_0({\bf m})}\right\},
\een
where  $\Delta_\ttt{A}=\sum_{i=1}^M (k_i-k_i')$ and $\Delta_\ttt{B}=\sum_{i=1}^M (m_i-m_i')$. This procedure generates a Markov chain, which, once converged, samples from the target distribution $\tilde{p}({\bf k},{\bf m})$. As discussed in the main text, the outlined probabilistic post-processing algorithm can also be considered as generalized post-selection: post-selection involves discarding results that do not match the corresponding conditions, while we deal with an acceptance-based random-walk on the space of detected events.

The probability distribution $\tilde{p}_0({\bf m})$ can be chosen, in principle, arbitrarily. For instance, one  may consider a regime where input states $\ket{{\bf m}}$ are uniformly distributed over a shell with a fixed photon number $\cal N$. This is in the spirit of scattershot boson sampling. In such a case (i.e., within this shell), $T(\{{\bf k'}, {\bf m'}\}|\{{\bf k}, {\bf m}\})={\rm min}\{1, \xi^{\Delta_\ttt{A}}\}$. Consequently, the average acceptance rate $\avg{T}$ is lower bounded as follows $\avg{T}= \avg{{\rm min}\{1, \xi^{\Delta_\ttt{A}}\}}\geq1/2$, since $\Delta_\ttt{A}$ can be seen as a random variable which takes non-negative (non-positive) values with probability 1/2. In other words, the average acceptance probability in this case is necessarily larger than 1/2, which ensures the fair convergence of our post-processing algorithm (similar arguments hold as well, e.g., for a thermal distribution over Fock states $\ket{{\bf m}}$). On the other hand, in a purely linear optics regime (e.g., in the limit $t_j=1, \forall j$) the detection of a single-photon pattern {\bf m} at the output of $\mathcal{U}_\ttt{B}$ can be seen as a random pattern of single photons input to the boson sampling circuit $\mathcal{U}_\ttt{A}\mathcal{U}_\ttt{B}^\ttt{T}$, yielding a detection of $N_\ttt{A}=N_\ttt{B}$ photons at its output (in general, however, the photon number in a Gaussian circuit is not conserved, i.e., $N_\ttt{A}\neq N_\ttt{B}$). Alternatively, if we desire to simulate the evolution of a specific fixed input state $\ket{{\bf m}_0}$, we choose $\tilde{p}_0({\bf m})=\prod_{i=1}^M \delta_{m_i,m_{0_i}}$, yielding $T(\{{\bf k'}, {\bf m'}\}|\{{\bf k}, {\bf m}\})={\rm min}\{1, \xi^{\Delta_\ttt{A}}\}$. The probability of detecting a tuple $\{{\bf k, m}_0\}$ at the output of $\mathcal{U}_\ttt{G}$ will be exponentially smaller than in the case of uniformly distributed tuples ${\bf m}$. Nevertheless, the technique remains valid. 

For the sake of completeness, we also present here the marginal probability $p({\bf m})$ of detecting a pattern {\bf m} of single photons at the output of the circuit $\mathcal{U}_\ttt{B}$:
\ben\nonumber
p({\bf m})=\sum_{k_1, \dots, k_M=0}^\infty\left|\bra{{\bf k}}\avg{{\bf m}|\psi_\ttt{out}}\right|^2=\avg{{\bf m}|\mathcal{W}_\ttt{B}\rho_\ttt{B}\mathcal{W}_\ttt{B}^\dagger|{\bf m}},
\een
where $\rho_{\ttt{B}}\equiv\ttt{Tr}_\ttt{A} \rho_{\ttt{in}}= \ttt{Tr}_\ttt{A} \ket{\psi_{\ttt{in}}}\bra{\psi_{\ttt{in}}}$ denotes the Gaussian state obtained after tracing out $\ket{\psi_{\ttt{in}}}$ over the modes entering the transformation $\mathcal{W}_\ttt{A}$. The state $\rho_{\ttt{B}}$ can be easily described in its phase-space representation using the formalism of Appendix~\ref{phase}. Namely, its covariance matrix reads
\begin{widetext}
\ben\nonumber
&& \sigma_{\ttt{B}}=\oplus_{j=1}^{M/2}\sigma_{\ttt{B}}^{(j)},\\
&&\sigma_{\ttt{B}}^{(j)}=\frac{1}{2}\begin{bmatrix}
	{\rm cosh\,} 2r      & 0  					      				& 0 										& \sqrt{1-t_j}\,{\rm sinh\,} 2r\\
	0                          & {\rm cosh\,} 2r      				& \sqrt{1-t_j}\,{\rm sinh\,} 2r & 0 \\
	0 				           & \sqrt{1-t_j}\,{\rm sinh\,} 2r  & {\rm cosh\,} 2r 	   				& 0\\
	\sqrt{1-t_j}\, {\rm sinh\,} 2r    &	0						& 0											 & {\rm cosh\,} 2r 
\end{bmatrix},
\een
\end{widetext}
where the parameter $r$ is defined such that ${\rm tanh}\, r\equiv \xi$. Given this covariance matrix, the corresponding probability $p({\bf m})$ reads~\cite{GBS},
\ben\label{a6}\nonumber
&&p({\bf m})=\frac{\ttt{Haf }A_{\bf m}}{m_1!\cdots m_M!\sqrt{\ttt{det}\, \tilde{\sigma}_{\ttt{out}}}}.
\een
In the above expression, $\tilde{\sigma}_{\ttt{out}}=1/2S_{\ttt{B}} \sigma_{\ttt{B}} S_{\ttt{B}}^\dagger$ ($S_{\ttt{B}}$ is the phase-space representation of the transformation $\mathcal{W}_\ttt{B}$), $A_{\bf m}$ is a $2 N_\ttt{B}\times 2 N_\ttt{B}$ matrix obtained from the matrix $A=\begin{bmatrix}0 & I_M  \\ I_M & 0 \end{bmatrix} \left[I_{2M}-\tilde{\sigma}_\ttt{out}^{-1}\right]$ by repeating $m_i$ times its $i$th  and $(i+M)$th columns and rows. Finally, $\tilde{\sigma}_{\ttt{out}}=\sigma_{\ttt{out}}+I_{2M}/2$ and the Hafnian of a $2K\times 2K$ matrix $X$ is defined as~\cite{hafnian}:
\be\label{a7}
\ttt{Haf}\,X=\sum_{\mu\in  \ttt{C}_{2K}} \prod_{j=1}^K X_{\mu(2j-1),\mu(2j)},
\ee
where C$_{2K}$ is the set of canonical permutations on $2K$ elements, obeying $\mu(2j-1)<\mu(2j)$ and $\mu(2j)<\mu[2(j+1)]$, $\forall j$. Although this probability is given in terms of a computationally hard matrix Hafnian, its evaluation is not required for implementing our Metropolised independent sampling algorithm as defined in the main text.

%This can be understood as follows. By summing Eq. (\ref{app3}) on $k$, we have
%\ben
%p({\bf m})= \sum_{k_1, \dots, k_M=0}^\infty A({\bf k},{\bf m}) \, \tilde{p}({\bf k}|{\bf m}) 
%\een
%If $A({\bf k},{\bf m})$ was independent of ${\bf k}$, then we would have $p({\bf m})= A({\bf m})$ and it would be necessary to compute Hafnians in order to run the simulation. Fortunately, the specific dependence of $A({\bf k},{\bf m})$ in ${\bf k}$ makes the algorithm practical.

Note also that in the limit $t_j=1$ ($\forall j$), $\sigma_{\ttt{B}}$ is the covariance matrix of $M$ thermal states and $p({\bf m})=(1-\xi^2)^M \xi^{2N}$, yielding, $p({\bf k},{\bf m})=\tilde{p}({\bf k},{\bf m})$. This regime corresponds to twofold scattershot boson sampling~\cite{twofold}. Indeed, if $t_j=1$ ($\forall j$), we have a set of $M/2$ two-mode squeezed vacuum states $\ket{\psi_{j}}$ injected into the circuits $\mathcal{U}_\ttt{G}$ in Fig. 2(a).

\section{Arbitrary Gaussian circuits}\label{proof2}

As already stated, the simulated circuit $\tilde{\mathcal{U}}_\ttt{G}$ represents a special instance of the Bloch-Messiah decomposition since the corresponding single-mode squeezers are equal by pairs $r^{(j_1)}=-r^{(j_2)}$, whereas in general, a Gaussian transformation necessitates a set of single-mode squeezers of arbitrary squeezing degrees. However, a slight modification of the $2M$-mode circuit $\mathcal{U}_\ttt{G}$ allows one to achieve any $M/2$-mode Gaussian transformation, i.e., at the expense of decreasing by half the number of its modes. We now construct the corresponding $2M$-mode linear-optical circuit $\mathcal{U}_\ttt{G}^{(2)}$ by slightly modifying $\mathcal{U}_\ttt{G}$. Namely, we replace the $M$-mode circuit $\mathcal{U}_\ttt{A}$ ($\mathcal{U}_\ttt{B}$) with two disjoint circuits $\mathcal{V}_\ttt{A}$ and $\mathcal{V}'_\ttt{A}$ ($\mathcal{V}_\ttt{B}$ and $\mathcal{V}'_\ttt{B}$), as illustrated in Fig.~\ref{figS1}(a). In turn, as opposed to $\tilde{\mathcal{U}}_\ttt{G}$ where all the modes entering $\mathcal{U}_\ttt{A}$ emerge from the preceding balanced beam splitters, in $\tilde{\mathcal{U}}_\ttt{G}^{(2)}$ we inject the upper output port of every balanced beam splitter to $\mathcal{V}_\ttt{A}$ while the lower one to $\mathcal{V}'_\ttt{A}$ (similarly for the modes entering $\mathcal{U}_\ttt{B}$). Next, we consider a pattern of photons ${\bf k}$ (${\bf k}'$) detected at the output of $\mathcal{V}_\ttt{A}$ ($\mathcal{V}'_\ttt{A}$) and a pattern ${\bf m}$ (${\bf m}'$) at the output of $\mathcal{V}_\ttt{B}$ ($\mathcal{V}'_\ttt{B}$). Additionally, we use notations $\sum_{i=1}^{M/2}k_i\equiv K_\ttt{A}$, $\sum_{i=1}^{M/2} k_i'\equiv K'_\ttt{A}$, $\sum_{i=1}^{M/2}m_i\equiv K_\ttt{B}$, $\sum_{i=1}^{M/2} m_i'\equiv K_\ttt{B}'$, $K_\ttt{A}+K_\ttt{B}\equiv K$ and $K_\ttt{A}'+K_\ttt{B}'\equiv K'$. 

Following the same time-unfolding formalism as in Sec.~\ref{time} and taking into account the relation between the two- and single-mode squeezers [inset in Fig.~2(b)], we find that the $2M$-mode circuit $\mathcal{U}_\ttt{G}^{(2)}$ is equivalent to two disjoint circuits $\tilde{\mathcal{U}'}_\ttt{G}^{(2)}$ and $\tilde{\mathcal{U}}_\ttt{G}^{(2)}$. Namely, $\tilde{\mathcal{U}'}_\ttt{G}^{(2)}$ and $\tilde{\mathcal{U}}_\ttt{G}^{(2)}$ are injected with the states $\ket{{\bf m}}$ and $\ket{{\bf m'}}$, respectively, and the respective single photon patterns ${\bf k}$ and ${\bf k}'$ are detected at their outputs. These circuits read:
\ben
&&\tilde{\mathcal{U}}_\ttt{G}^{(2)}=  \mathcal{V}_\ttt{A}  \left[\otimes_{j_1=1}^{M/2} \mathcal{U}_\ttt{SS}^{(j_1)}\right]  \mathcal{V}_\ttt{B}^{\ttt{T}}  ,\\
&&\tilde{\mathcal{U}'}_\ttt{G}^{(2)}= \mathcal{V'}_\ttt{A}  \left[\otimes_{j_2=1}^{M/2} \mathcal{U}_\ttt{SS}^{(j_2)}\right]  \mathcal{V'}_\ttt{B}^{\ttt{T}},
\een
where $\mathcal{U}_\ttt{SS}^{(j_1)}$ and $\mathcal{U}_\ttt{SS}^{(j_2)}$ differ by the sign of their squeezing degrees for every pair $j_1=j_2$. Importantly, each of the above equations represents the Bloch-Messiah decomposition of an $M/2$-mode Gaussian transformation with no restrictions. In other words, by means of a $2M$-mode linear-optical circuit $\mathcal{U}_\ttt{G}^{(2)}$ injected with $M$ TMSs, one can simulate sampling from the joint input and output distributions of any target $M/2$-mode Gaussian transformation by means of its Bloch-Messiah decomposition, which, in turn, is realized via $\tilde{\mathcal{U}}_\ttt{G}^{(2)}$. Although we decrease by half the number of available modes, we are able to implement two Gaussian transformation simultaneously. That is, we can choose the pairs $\{\mathcal{V}_\ttt{A}, \mathcal{V}_\ttt{B}\}$ and $\{\mathcal{V}'_\ttt{A}, \mathcal{V}'_\ttt{B}\}$ of two linear-optical circuits independently. 

Following the same reasoning as in Appendix~\ref{proof}, the joint photon-counting probability distribution $p({\bf k, k', m, m'})$ for the circuit $\mathcal{U}_\ttt{G}^{(2)}$ can be written down as
%\begin{widetext}
\ben\label{app4}
&&p({\bf k},  {\bf m}, {\bf k'},  {\bf m'})=\left|\bra{{\bf k, k'}} \avg{{\bf m, m'}|\mathcal{U}_\ttt{G}^{(2)}|\psi_\ttt{in}}\right|^2=\\ \nonumber
&&\frac{(1-\xi^2)^M \xi^{2N}}{\prod_{j=1}^{M/2}{t_j}}
\left|\bra{\bf k}\tilde{\mathcal{U}}_\ttt{G}^{(2)}\ket{\bf m}\right|^2 
\left|\bra{\bf k'}\tilde{\mathcal{U}}_\ttt{G}'^{(2)}\ket{\bf m'}\right|^2\\ \nonumber
&&\equiv A({\bf k},{\bf m}, {\bf k'},  {\bf m'})\tilde{p}({\bf k}|{\bf m})\tilde{p}'({\bf k'}|{\bf m'}).
\een
%\end{widetext}
Consequently, a random-walk sampling algorithm can be adapted here, analogous to the case of the circuit $\mathcal{U}_\ttt{G}$, in order to simulate sampling from the probability distribution $\tilde{p}({\bf k},{\bf m})\tilde{p}'({\bf k'},{\bf m'})=\tilde{p}_0({\bf m, m'})\tilde{p}({\bf k}|{\bf m})\tilde{p}'({\bf k'},{\bf m'})$, with a beforehand chosen distribution $\tilde{p}_0({\bf m, m'})$ of input states $\ket{{\bf m, m'}}$.

Finally, it is worth nothing that Gaussian circuits $\tilde{\mathcal{U}}_\ttt{G}$ and $\tilde{\mathcal{U}}_\ttt{G}^{(2)}$ do not conserve the number of photons. However, if photon detection happens immediately after the row of beam splitters $\mathcal{U}_\ttt{BS}^{(j)}$, the resulting time-unfolded circuit, both for $\tilde{\mathcal{U}}_\ttt{G}$ and $\tilde{\mathcal{U}}_\ttt{G}^{(2)}$, corresponds to a set of $M$ disjoint two-mode squeezers. For a two-mode squeezer, the photon number difference at its input is equal to the photon number difference at its output. That is, $m_i-m_{i+1}=k_i-k_{i+1}$ ($i=1, \dots, M$).

%\section{Extras}
%Let us recall the phase-space description of a $M$-mode Gaussian transformation in terms of a symplectic matrix $S$~\cite{footnote1}. The latter maps the vector of input mode operators $\hat{R}\equiv\{\hat{a}_1, \dots, \hat{a}_M, \hat{a}_1^\dagger, \dots, \hat{a}_M^\dagger\}$ onto the vector of output mode operators $\hat{Q}\equiv\{\hat{b}_1, \dots, \hat{b}_M, \hat{b}_1^\dagger, \dots, \hat{b}_M^\dagger\}$, namely~\cite{suppl}
%\begin{equation}\label{1}
%\hat{Q}_k=\sum_{l=1}^{2M}  S_{{kl}} \, \hat{R}_l.
%\end{equation}
%n particular, a beam-splitter transformation of transmissivity $t_j$ is defined by the symplectic matrix
%\ben
%&& S_\ttt{BS}^{(j)}=\begin{bmatrix}
%	R_\ttt{BS}^{(j)} & 0  \\
%	0 & R_\ttt{BS}^{{(j)}}
%\end{bmatrix},
%\een
%with 
%\ben
%R_\ttt{BS}^{(j)}=\begin{bmatrix}
%	\sqrt{t_j} & \sqrt{1-t_j}  \\
%	-\sqrt{1-t_j} & \sqrt{t_j} 
%\end{bmatrix}.
%\een
%being a rotation matrix. For simplicity, we shall use $S_{\ttt{BS}}$ to designate a balanced beam-splitter transformation ($t=1/2$).
%In turn, the phase-space description of a two-mode squeezer of gain $g_j$ is given by the matrix
%\ben
%&& S_\ttt{TS}^{(j)}=\begin{bmatrix}
%	\sqrt{g_j} & 0  & 0 & \sqrt{g_j-1}\\
%	0 & \sqrt{g_j} & \sqrt{g_j-1}& 0 \\
%	0 & \sqrt{g_j-1}  & \sqrt{g_j} & 0\\
%	\sqrt{g_j-1} & 0 & 0& \sqrt{g_j}
%\end{bmatrix}.
%\ee

\end{document}